\documentclass[12pt,preprint]{aastex}

\input{epsf}

\newcommand{\be}{\begin{equation}}
\newcommand{\ee}{\end{equation}}

\usepackage{graphicx}
\usepackage{txfonts}
\usepackage{natbib}
\usepackage[usenames]{color}

\begin{document}

\shorttitle{Multi-messenger model for M82}

\title{Multi-messenger model for the starburst galaxy M82 }

\author{Elsa de Cea del Pozo\altaffilmark{1},
Diego F. Torres\altaffilmark{1,2},
Ana Y. Rodriguez Marrero\altaffilmark{1} }

\altaffiltext{1}{Institut de Ciencies de l'Espai (IEEC-CSIC)
  Campus UAB, Fac. de Ciencies, Torre C5, parell, 2a planta, 08193
  Barcelona,  Spain. E-mail: decea@ieec.uab.es, arodrig@ieec.uab.es}
\altaffiltext{2}  {Instituci\'o Catalana de Recerca i Estudis Avan\c{c}ats (ICREA), Spain. E-mail: dtorres@ieec.uab.es }

\begin{abstract}
In this paper, a consistent model of the multifrequency emission of the starburst galaxy M82, from radio to gamma-rays is presented and discussed. Predictions for observations with Fermi, MAGIC II/VERITAS and CTA telescopes are made. 
The model is also used to self-consistenty compute the (all flavors) emission of neutrinos resulting from this starburst galaxy, what can be used in considerations of the diffuse contributions of such objects.

\end{abstract}

\begin{keywords}
{Starburst galaxy (individual M82), $\gamma$-rays: theory, $\gamma$-rays: observations}
\end{keywords}

\section{Introduction}
\label{intro}

M82 is a near starburst galaxy. It can be seen nearly edge-on (77$^\circ$, Mayya et al. 2005) and has a gas content mostly concentrated in the inner 2 kpc. This galaxy presents a high luminosity both in the far infrared (IR) and X-ray domain (10$^{44}$ erg s$^{-1}$ and 10$^{40}$ erg s$^{-1}$ respectively (e.g., Ranalli et al. 2008).
As part of the M81 group, M82 shows hints of an encounter with some of its members 1 Gyr ago, specifically with the dwarf galaxy NGC 3077 and the former M81, materialized in an intergalactic gas bridge of 20 kpc. Up to 10 kpc above the plane of M82, galactic superwinds can be detected. Following HI streamers, the external part of the disk ($>$5 kpc) presents a warped form. On the other hand, the inner part (300 pc) harbours a starburst. Around this region, a molecular ring of 400 pc radius and a near-IR bar of $\sim$ 1 kpc length can be detected {(Telesco et al. 1991)}. From the tips of this bar, two symmetrical arms emerge (Mayya et al. 2005). This led to a change in the morphological classification. Previously, an irregular shape was assumed due to optical appearance: bright star-forming knots interspersed by dusty filaments { (O'Connell et al. 1978)}; now it seems more likely that M82 is a SBc galaxy. Freedman et al. (1994) established a distance for M81 of 3.63 $\pm$ 0.34 Mpc, thanks to the discovery of new Cepheids by the Hubble Space Telescope (HST) in this galaxy. A few years later, Sakai \& Madore  (1999) found a distance of 3.9 ($\pm0.3)_{random}$($\pm0.3)_{systematic}$ Mpc for M82, based on the detection of the red giant branch stars using HST photometry. Both distances are consistent within errors, the latter is the one used in this work. The starburst region is located in the inner part of the galaxy (radius of $\sim$300 pc, height of  $\sim$200 pc, e.g., see V\"olk et al. 1996 and references therein, and Mayya et al. 2006), meanwhile the rest of the emission extends to a thin disk up to 7 kpc (see e.g., Persic et al. 2008, and references therein). A significant amount of molecular material has been found in the central region, where most of the starburst activity (see below the discussion on the supernova explosion rate) is located. 
The content of the gas mass in the whole galaxy range from $2.9 \times 10^9 M_\odot$ (HI) by Crutcher et al (1978) and $2.9 \times 10^8 M_\odot$ ($H_2$) by Young \& Scoville (1984) to the more recent results found by Casasola et al. (2004) $7 \times 10^8 M_\odot$ (HI) and $1.8 \times 10^9 M_\odot$ ($H_2$), where different assumptions for distance and the normalization  of luminosity were made.  
Weiss et al. (2001) report for the starburst region around $2 \times 10^8 M_\odot$ ($H_2$), using separate methods for the determination of CO and $H_2$ densities 
(see this paper for further details). This value essentially agrees with those estimated from 450 $\mu$m dust continuum measurements (Smith et al. 1991) and from CO$(2\rightarrow1)$ intensities (Wild et al. 1992), and is, therefore, the one used in the determination of the uniform density for the model of this paper ($\sim 180$ cm$^{-3}$).

With the advent of Fermi, the upgrade of MAGIC to an stereoscopic observatory, and with VERITAS on-line, it is important to have the most detailed model of M82  to be able to feedback from observations our knowledge of the cosmic-ray population and the physical environment of nearby starbursts. In addition, with the recent on-going discussion about the starbursts' contribution to the neutrino background, it is important to have consistent and detailed models of neutrino production in specific galaxies, where the assumptions made in previous studies could be studied in detail and observational detectability with e.g., km$^3$-observatories could be assessed.


\section{Theoretical approach}
\label{theory}

\begin{figure} [t]
\centering
\includegraphics[width=.95\columnwidth,trim=0 5 0 10]{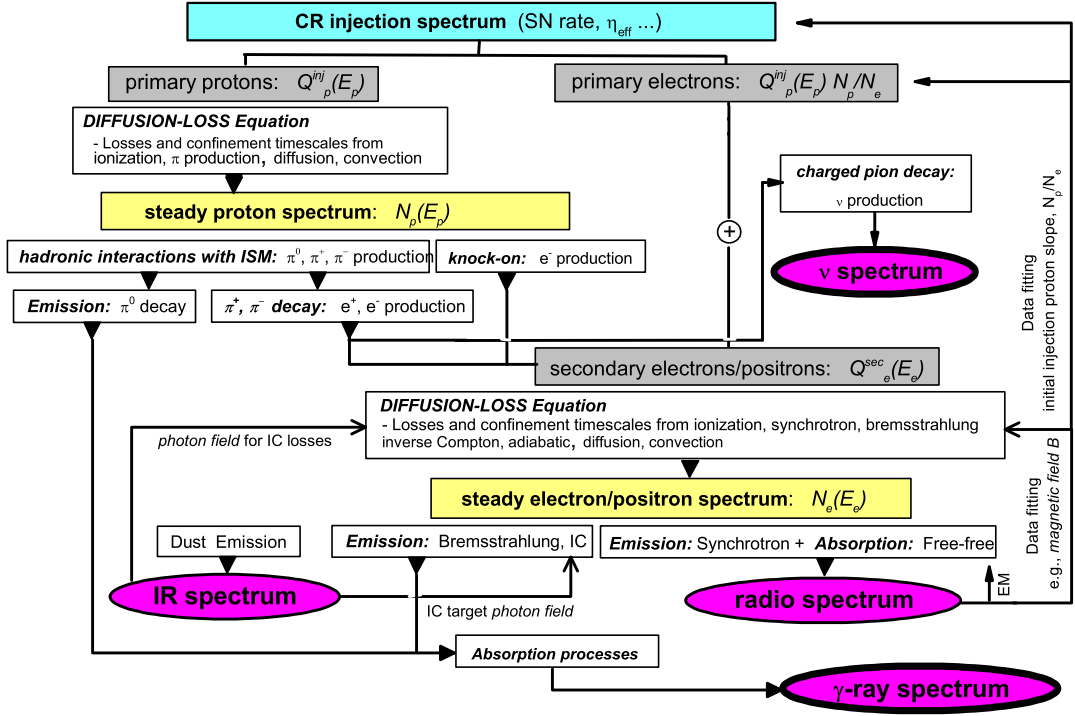}
\caption{Code flow of ${\cal Q}$-{\sc diffuse}, adapted from Torres and Domingo-Santamar\'ia (2005) to take into account new updates developed for this work. }
 \label{flow}
\end{figure}

The aim of this study is to obtain multi-frequency/multi-messenger predictions for the photon and neutrino emission coming from the central region of M82, where the inner starburst is located. We seek a consistent model, where the different components of the emission (from radio to TeV photons and neutrinos) can all be tracked to one and the same original cosmic-ray population, and result as a consequence of all electromagnetic and hadronic channels from the primary and subsequently-produced secondary particles. With this model at hand, we explore a range of uncertainties in the parameters, which may correspondingly lift up or lower the high energy end of the spectrum. 
In order to perform such a study we have updated the code  ${\cal Q}$-{\sc diffuse} (Torres 2004) in several ways. ${\cal Q}$-{\sc diffuse} solves the diffusion-loss equation for electrons and protons and finds the steady state distribution
for these particles subject to a complete set of losses in the
interstellar medium (ISM). Subsequently, it computes secondaries from hadronic
interactions (neutral and charged pions) and Coulomb processes
(electrons), and gives account of the radiation or decay products
that these particles produce. Secondary particles (photons, muons,
neutrinos, electrons, and positrons) that are in turn produced by
pion decay  are calculated too. Additional pieces of the code compute the dust
emissivity, and the IR-FIR photon density, which is consistently used both as
target for inverse Compton scattering and to model the radiation at
lower frequencies. For radio photons, we
compute, using the steady distribution of electrons, 
synchrotron radiation, free-free emission, and absorption. Finally, opacities to $\gamma \gamma$ and $\gamma
Z$ processes are computed, as well as absorbed $\gamma$-ray fluxes,
using the radiation transport equation. Our implementation of ${\cal Q}$-{\sc diffuse} includes several upgrades. Some are strictly technical, allowing for a more versatile and automatic input-output interaction; some are physical, we have developed neutrino-production subroutines for all neutrino channels in the decay of positively and negatively charged pions and make use of a recent parameterization of $pp$ interactions by Kelner et al. (2006). The flow of the ${\cal Q}$-{\sc diffuse} code is shown in Fig. \ref{flow}. Several loops can be distinguished there, they are used in order to determine a self-consistent set of parameters such that all predictions for the different bands of the electromagnetic spectrum are consistent with data. 

Previous studies of diffuse high energy emission, and of electron
and positron production, with different levels of detail and aims,
go back to the early years of $\gamma$-ray astronomy. A summary of
these first efforts is summarized by Fazio (1967),
Ginzburg and Syrovatskii (1968), 
and Ramaty \& Lingenfelter (1968), then followed by Maraschi et al.
(1968), and Stecker (1977), among many others. 
Secondary particle
computations have a similarly long history
see, e.g., Stecker (1969), Orth and Buffington (1976). Relatively more recent efforts include
Drury
et al. (1994),
Moskalenko \& Strong (1998), Strong \& Moskalenko (1998), Markoff et
al. (1999), and Fatuzzo \& Melia (2003). Here, the general ideas used by
Paglione et al. (1996) and Blom et al. (1999), when modelling nearby
starbursts galaxies, are followed. These, in turn, closely track
Brown \& Marscher's (1977) and Marscher \& Brown's (1978), regarding
their studies of close molecular clouds.

We emphasize here that this approach lacks detail of modeling of single sources, we deal not with modeling isolated SN explosions or SN remnants, and not with
the issues of them exploding within the wind bubbles of their progenitors and the concurrent losses in these environments, but rather on average properties of the whole starburst region, seen as a CR injector. To mix these two aims would require an effort that is beyond both, current observational capabilities of what is known in M82, as well 
as beyond what is possible for this work: essentially, to make quantitative estimates in this way one would need to understand the features of each and all the SN explosions in M82, model them in detail individually,  
and from that build (i.e., from an stacking of a source-by-source modeling) an averaged scenario. This has not been achieved even for our own Galaxy. Similarly, our approach to the modeling of the super-wind is also simple in order to encompass it with the rest of the components of the model, see Everett et al. (2008) for a more detailed analysis of the properties of the Galactic super wind, see also Gallagher \& Smith (2005).

The assumption of a uniform distribution of accelerators in the inner region of the starburst justifies to have (and compute it via solving the full diffusion-loss equation) the existence of a steady state distribution therein.  
The limited size and compactness of the starburst region is compatible with an steady state electron distribution in that central region. 
The  { steady state particle distribution} is computed as the result of an injection
distribution being subject to losses and secondary production in
the ISM. At sufficiently high energies, the
injection proton emissivity is assumed to
have the following form as a function of proton kinetic energies $
Q_{\rm inj}(E_{\rm p,\, kin}) = K ({E_{\rm p,\, kin}}/{\rm
GeV})^{-p} \exp(- E_{\rm p,\, kin} / E_{\rm p, \, cut})$, where $p$ is a power-law index, $K$ is a normalization constant, $E_{\rm p, \,cut}$ is an energy cutoff in the accelerated particles injected (assumed as 100 TeV, with a range explored below). Units are
$[Q]$= {\rm GeV}$^{-1}$ {\rm cm}$^{-3}$  {\rm s}$^{-1}$.
To get a numerical prior for the normalization $K$ is assumed to come from the total power transferred
by supernovae into CRs kinetic energy within a given volume; thus, the supernova rate is essential to fix how high is the cosmic-ray sea $
\int_{E_{\rm p,\,kin,\, min}}^{E_{\rm p,\, kin,\, max}} Q_{\rm
inj}(E_{\rm p,\, kin}) E_{\rm p,\, kin} dE_{\rm p,\, kin}  
 \equiv {  \eta {\cal P} {\cal R} }/ {V}.
$ 
${\cal R}$ is the rate of supernova explosions, $V$ being its volume, and $\eta$, the
transferred fraction of the supernova explosion power (${\cal P}
\sim 10^{51}$ erg) into CRs. The average rate of power transfer is assumed to be 10\%.
Our assumption of the $p$ value is not a-priori, it is rather an a posteriori choice made by what it seems to be a slightly better fit to all multi-wavelength data and acts as an average description of the injected cosmic-ray sea in the star forming region, where multiple nearby shocks could contribute.

As noted by Domingo-Santamar\'ia and Torres (2005), at low energies the distribution of cosmic rays is probably flatter, e.g., it would be given by equation (6) of Bell (1978), correspondingly normalized. We have numerically verified that to neglect this difference at low energy does not  produce any important change in the computation of secondaries, and especially on $\gamma$-rays at the energies of interest.  In here, we  assume, e.g, as in  Berezhko et al. (2006) that
electrons are injected into the acceleration process at the shock fronts. We 
choose the electron injection rate such that the electron-proton 
ratio  
is a constant to be determined from the synchrotron 
observations. The diffusive transport equation takes care of the changes produced after injection onto this distribution.

The diffusion-loss equation is given by (see, e.g.,
Longair 1994, p. 279; Ginzburg \& Syrovatskii 1964, p. 296) $ -
D \bigtriangledown ^2 N(E)+ {N(E)} / {\tau(E)} - {d}/{dE}
\left[ b(E) N(E) \right] - Q(E) = - {\partial N(E)} / {\partial
t} \label{DL} $. In this equation, $D$ is the scalar diffusion
coefficient, $Q(E)$ represents the source term appropriate to the
production of particles with energy $E$, $\tau(E)$ stands for the
confinement timescale, $N(E)$ is the distribution of particles
with energies in the range $E$ and $E+dE$ per unit volume, and
$b(E)=-\left( {dE}/{dt} \right)$ is the rate of loss of energy.
The functions $b(E)$, $\tau(E)$, and $Q(E)$ depend on the kind of
particles considered. In the steady state,  $ {\partial N(E)}/{\partial t}
=0,$ and when the spatial dependence is considered to be irrelevant $ D \bigtriangledown ^2 N(E) =0.$  Then the diffusion-loss equation can be
numerically solved by using the corresponding Green function $G$,  such that for any given source function, or
emissivity, $Q(E)$, the solution is $ N(E) = \int_E^{E_{\rm
max}} d{E^\prime} Q({E^\prime}) G(E,{E^\prime}). \label{SOL-DL} $
%
%
The confinement
timescale will  take into account that particles can be diffusing away, can be
carried away by the collective effect of stellar winds and
supernovae,  or can be affected by pion production (for CRs). 
Pion
losses (which are catastrophic, since the inelasticity of the
collision is about 50\%) produce a loss timescale $\tau_{\rm
pp}^{-1}=(dE/dt)^{\rm pion}/E$ (see, e.g., Mannheim \&
Schlickeiser 1994). Thus, in general, for energies higher than the
pion production threshold $ \tau^{-1}(E)= \tau_D^{-1}
 + {\tau_c}^{-1} +
\tau_{\rm pp}^{-1}. \label{T-P} $ 
$\tau_c$, the convective timescale, is $\sim R / V$,
where $V$ is the collective wind velocity. Strickland (1997) found a value of 600 km s$^{-1}$ for the wind velocity, and we use this value below. However, we studied the response of the model to an arbitrary doubling or halving of this value and found that the uncertainty so-introduced is smaller than those produced by other parameters.
Finally, assuming an
 homogeneous distribution of accelerators in the central
 hundreds of parsecs of M82, the characteristic escape
time is that from the homogeneous diffusion model (Berezinskii et al. 1990,
p. 50-52 and 78) $ \tau_D= {R^2}/( {2D(E)}) ={\tau_0}/( \beta
(E/{\rm GeV})^{\mu} ), \label{T-P0} $ where $\beta$ is the
velocity of the particle in units of $c$, $R$ is the spatial
extent of the region from where particles diffuse away, and $D(E)$
is the energy-dependent diffusion coefficient, whose dependence is
assumed $\propto E^{\mu}$, with $\mu \sim 0.5$ and $\tau_0$ is the
characteristic diffusive escape time at $\sim$ 1 GeV.  Results that follow use ${\tau_0}=10$ Myr, but a range was explored to judge uncertainties.
%
%
%
%
 For electrons, the total rate
of energy loss considered is given by the sum of that involving
ionization, inverse Compton scattering, adiabatic, bremsstrahlung, and
synchrotron radiation. 
The Klein-Nishina cross section is used.

\begin{figure*} 
\centering
\includegraphics[width=.48\columnwidth,trim=0 5 0 10]{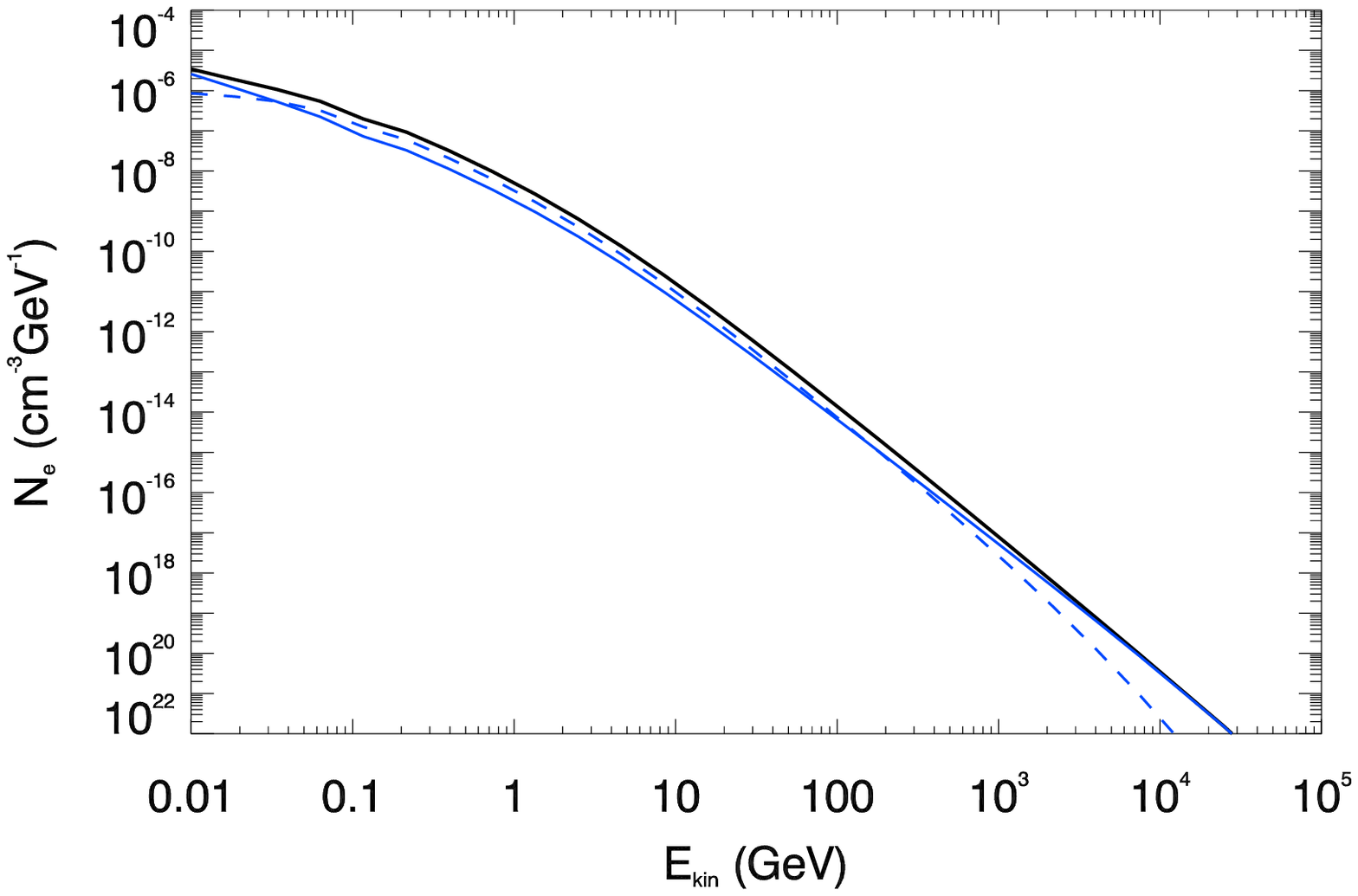}
\includegraphics[width=.48\columnwidth,trim=0 5 0 10]{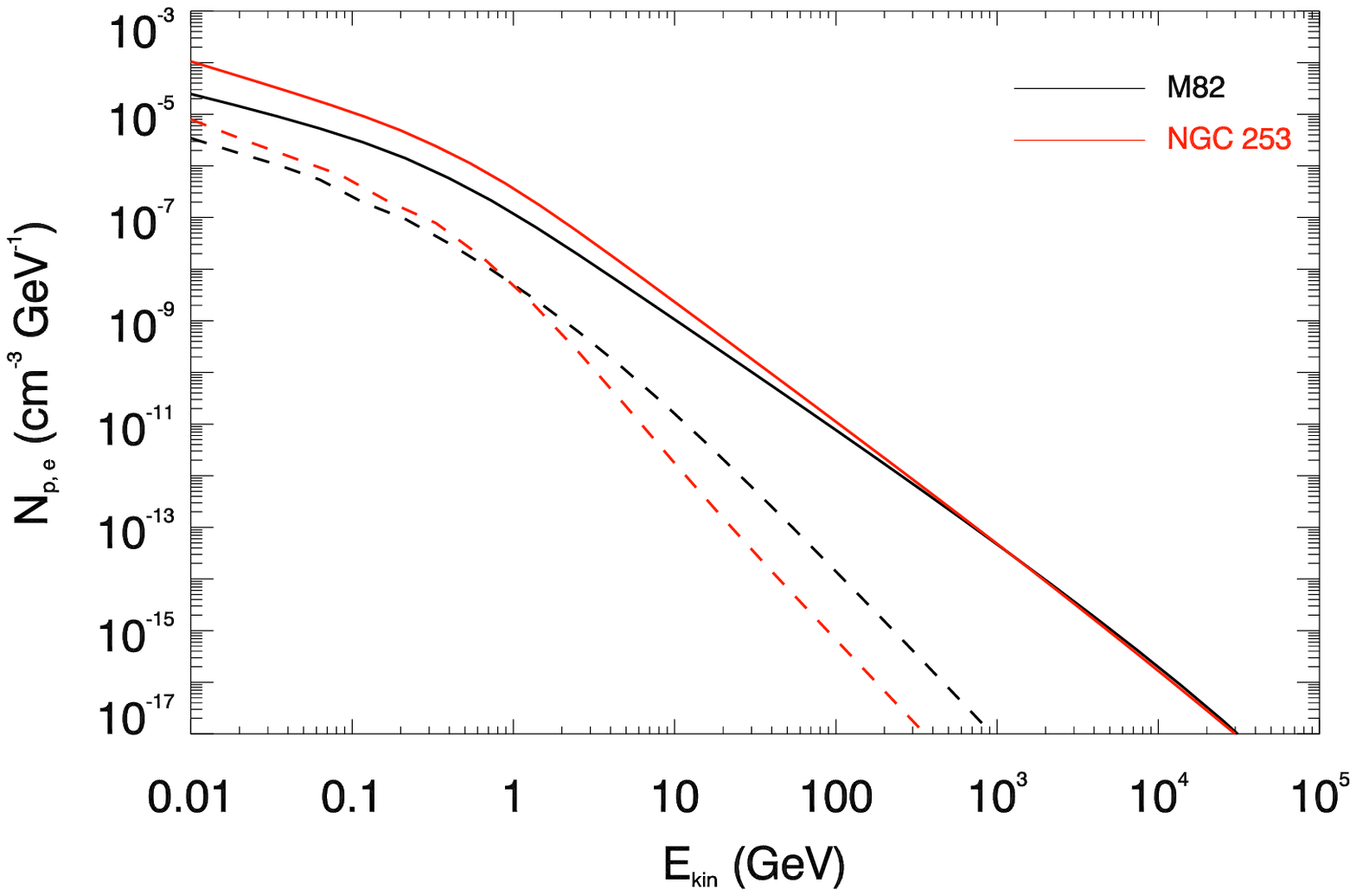}
\caption{Left: Comparison of the steady population of electrons that would result after solving the diffuse loss equation injecting only primary (solid blue) and only secondary (dashed blue) particles (from knock-on and pion decay) in the inner region of M82. The total steady electron population (resulting from the injection of both primary and secondary electrons is also shown, solid black). Parameters used in this figure correspond to those with the model which overall results are presented below (Fig. \ref{Fig2}--\ref{Fig5}). Right: Steady proton (continuous line) and electron (dashed) distributions in the innermost region of both M82 (black) and NGC 253 (red).
}
 \label{comp}
\end{figure*}

A natural comparison can be made between this approach and previous studies of the central starburst region of M82, in particular, Aky\"uz et al. (1991), V\"olk et al. (1996), Paglione et al. (1996), and Persic et al. (2008). 
The earlier work made by Aky\"uz et al. focused on $\gamma$-ray emission at the GeV regime (integral flux above 100 MeV) and basically provides a simple order of magnitude estimation with data available at that time. V\"olk et al. also did not 
provide a multi-frequency model, but rather consider only neutral pion decay from protons to obtain an order of magnitude estimation of the flux at high and very-high gamma-rays. This approach was also taken (with minor differences) by Pavlidou and Fields (2001) and Torres et al. (2004). These computations (having a different aim than the one here, which is providing a consistent model all along the electromagnetic spectrum) do not  include a computation of secondaries and their emission at low energies.   
The work by Paglione et al. (1996) is more complex, and it was discussed in detail, in particular, when compared with results by ${\cal Q}$-{\sc diffuse} on NGC 253, by Torres (2004) and Domingo-Santamar\'ia and Torres (2005), where we refer the reader.
There are several different physical and methodological considerations embedded in their model than the ones assumed here, in addition of not providing high-energy (TeV) predictions. None of this work consider neutrino emission.
The most recent study on M82, from Persic et al. (2008) also differs from ours in some key parameters and, most importantly, in the method and assumptions for the modeling. 
Among the physical parameters which values are given by Persic et al., the distance (3.6 Mpc, which is actually the one to M81), the dust emissivity index ($\sigma = 1$), the proton to electron primary ratio ($N_p/N_e=200$), the magnetic field ($180\mu G$) and the slope of primary injection spectrum (2.4) are all different from the ones used or derived in the present study. The method by which these (latter) values are fixed by Persic et al. differ from ours. For instance, Persic et al. obtained the magnetic field by assuming equipartition with primary particles; while we find the magnetic field by a multi-frequency analysis that takes into account also secondary particles produced: As we show in Fig. \ref{comp}, left panel, and was found also in the cases of other highly-dense environments,  the steady population of electrons that would result after solving the diffusion loss equation injecting only primary and only secondary  (i.e., those electrons and positrons resulting from knock-on and pion decay in the inner region of M82) particles are comparable, so that
it is not a safe  assumption to consider that the primary population dominates to compute the equipartition field. In the numerical treatment by Persic et al. (2008), the path is thus inverse to ours. They seemingly fix the normalization of the proton spectrum starting from an assumed $N_p/N_e$ factor from equipartition, with the normalization for electrons $N_e$ in turn obtained from the radio spectrum. In our case, we start from protons, with the supernova explosions injecting primaries which are left to interact by all possible process producing electrons which in turn radiate synchrotron. It is this (not at all negligible) secondary electrons contribution what is taken into account when computing $B$ and $N_p/N_e$ in order for the data to be consistently matched at all frequencies (see Fig. \ref{flow} for further clarification). Neither equipartition is a priori assumed, nor the primary electron population is a priori fixed from the radio spectrum in our approach. Finally, neutrino emission is scaled from gamma-ray fluxes by Persic et al., while computed by use of correspondingly parameterized cross sections up to tertiary particles interactions in this work.

\section{Results and Discussion}
\label{result}

The steady population of protons and electrons of the starburst galaxy M82 which results from our model is shown in Fig. \ref{comp}, right, where a power-law index $p = 2.1$ for the injection of primary relativistic hadrons is used, and to fix numerics, an exponential cutoff is set to 100 TeV (no significant difference is observed even at energies as high as 10 TeV if it is changed to half this value). A range of values is studied for the normalization factor for the hadronic injection spectrum too, as it depends on other sensitive parameters, subject to uncertainties. For instance, the rate of SN explosion per year was assumed to be 0.3 SN yr$^{-1}$ in earlier works, but recently it is more commonly referred to as  $\sim$0.1 SN yr$^{-1}$.
In the present work, both a high and a low value for the SN rate were studied, revealing a range of uncertainties, as can be seen in Fig. \ref{Fig5}, where we have also consider other exponential cutoffs. 

Indeed, the discussion about the supernovae explosion rate is on-going. The highest value of 0.3 SN yr$^{-1}$ is used by, e.g., V\"olk et al. (1996) and other authors thereafter, and it is based on  Kronberg et al. (1985); which compiles different estimations, starting from 0.1 SN yr$^{-1}$ by Kronberg \& Wilkinson (1975), based on the total non-thermal emission, going up to 0.16  yr$^{-1}$ (based only in IR excess), then
0.2-0.3 by Kronberg \& Sramek (1985) based on direct monitoring of variability of discrete sources,  
and still going up to 0.3 SN yr$^{-1}$ Rieke et al. (1980), based on estimating the number of new radio sources.
But critics to Kronberg \& Sramek (1985) and in general to high values of the supernova remnant rate have arised based mainly in that the rate of detection of new radio sources does not correspond to them (e.g., see de Grijs 2001, Mc Leod et al. 1993, Bartel et al. 1987). Lower values for the rates (0.07-0.08 SN yr$^{-1}$) have also been recently favored by Fenech et al. (2008) who made an 8-days deep MERLIN radio imaging of SNR in the starburst and use two different methods (on one side, they assumed that the SNRs in M82 that are more radio luminous than Cas A are younger than it; on the other, assuming that the SNR are in free expansion and using the measured $N(<D) - D$ plot) to compute the explosion rate. 

\begin{figure}
\centering
\includegraphics[width=0.475\columnwidth,trim=0 5 0 10]{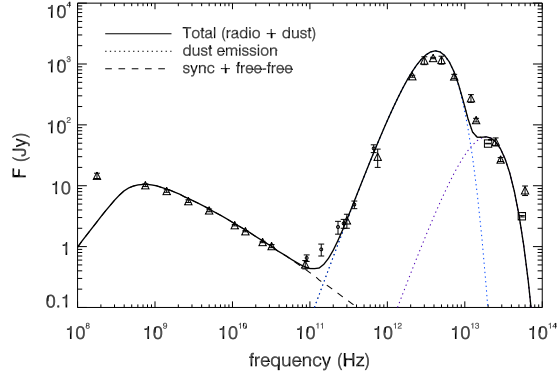}
\caption{Multi-frequency spectrum of M82 from radio to infrared. The observational data points correspond to: Klein et al. (triangles, 1988), Hughes et al. (circles, 1994) and Forster  et al. (squares, 2003) (and references therein in each case). The results from modelling correspond to: synchrotron plus free-free emision (dashed), dust emission (dotted) splitted in a cool (blue, T$_{c} = 45 K$) and a warm (purple,T$_{w} \simeq 200 K$) component, and the total emission from radio and IR emission (solid).}
 \label{Fig2}
\end{figure}

\begin{figure*}
\centering
\includegraphics[width=0.48\columnwidth,trim=0 5 0 10]{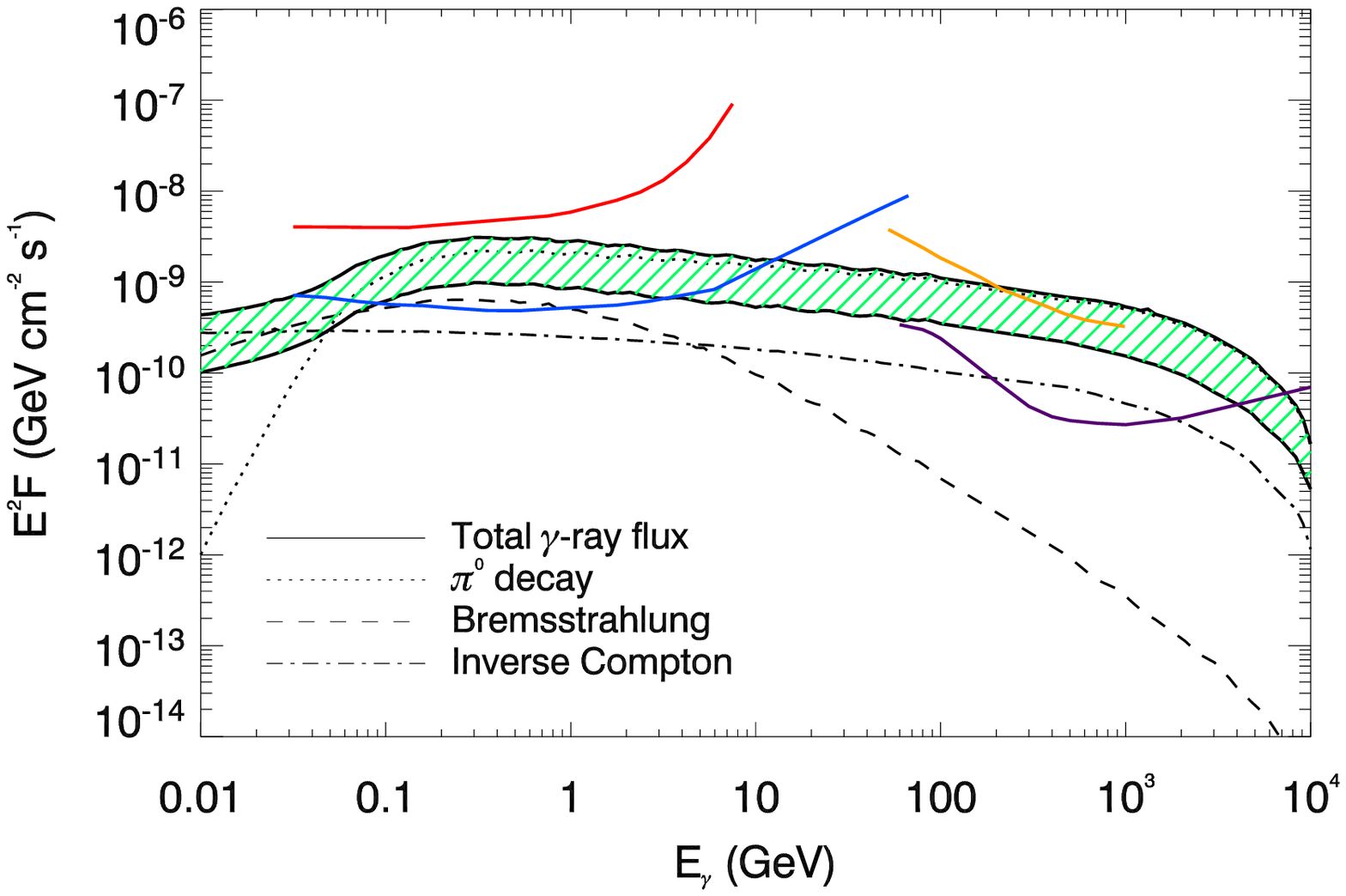}
\includegraphics[width=.5\columnwidth,trim=0 5 0 10]{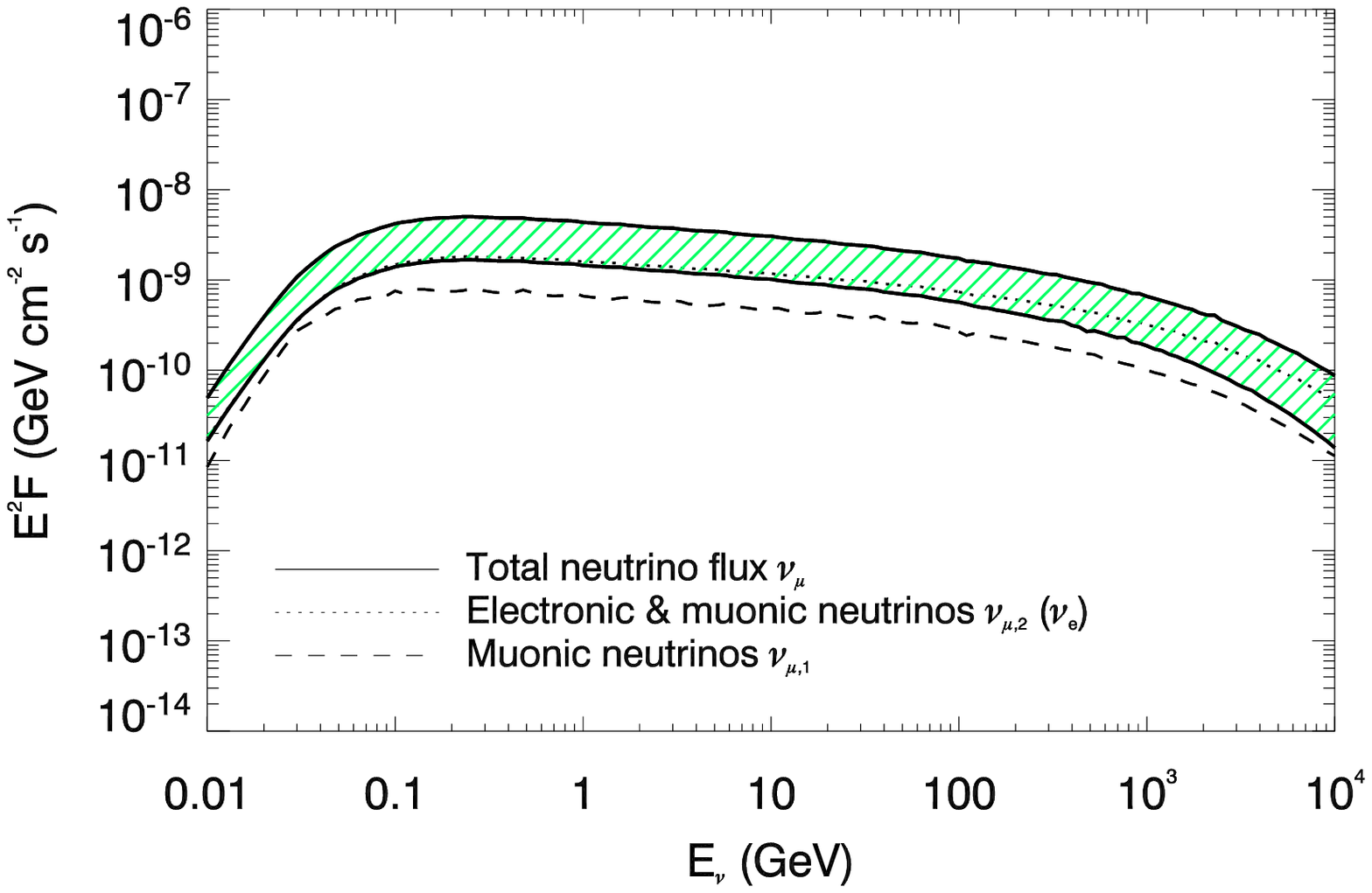}
\caption{Left: Energy distribution of the differential gamma-ray fluxes, exploring a range of uncertainties in supernova explosion rate and cutoffs in the primary energy, as it is explained in the text. The sensitivity curves for EGRET (red), Fermi (blue), MAGIC (purple), all from Funk et al. 2008, and the intended one for the forthcoming Cerenkov Telescope Array (CTA, violet) are shown. Right: 
Differential neutrino flux predictions from the inner region of M82, total and separated in different channels. The neutrino predictions make use of the same explored parameters already presented in the left diagram and explained in the text.}
 \label{Fig5}
\end{figure*}

The primary electron population is taken to be proportional to the injection proton spectrum, scaled by the ratio between protons and electrons, $N_p/N_e$. The value is set later on, together with others, e.g., the magnetic field, in a recursive application of our model, as a fit to radio data. 
Together with the primary electron population, all secondary  electrons produced are taken into account when computing the steady state. As a comparison, in Fig. \ref{comp}, right, we also plot the steady particle populations present in NGC 253 (from Domingo-Santamar\'ia and Torres 2005). Both curves present a similar behavior and level than those corresponding to M82, emphasizing the similaritity of the initial characteristics of these two starbursts as possible high-energy sources.
With the steady electron population determined, we proceed to compute a multi-frequency spectrum from radio up, and compare with experimental data. 

Fig. \ref{Fig2} shows the best match from our model, a multifrequency spectrum is overplotted to previous radio and IR data (see caption for details). At the lowest frequencies, radio data is no longer reduced to the central starburst region. Due to angular resolution of the observing instrument, emission at this frequency comes from the whole galaxy and cannot be separated from that coming only from the inner region, so our model is  just a component of the total radiation below a few $10^8$ Hz. Along the part of the spectrum in which synchrotron emission dominates, several parameters can be determined through repeated iterations, e.g., magnetic field, EM, slope of proton injection, and $N_{p}/N_{e}$ ratio.
In the range of uncertainty mentioned above, the lowest value for the magnetic field, $120 \mu G$, corresponds to the highest steady distribution (i.e., with the highest supernovae explosion rate assumed, and a value $N_{p}/N_{e}$=30). Inversely, a low steady distribution implies the need for greater losses by synchrotron in order to match the data, so leading the magnetic field up to $290 \mu G$. The general fit to observations is quite good in each case. These values of magnetic field are in agreement with previous results in this (V\"olk et al. 1989) and are also similar to the ones found for the disk of
Arp 220 (Torres 2004) or NGC 253 (Domingo Santamaria and Torres 2005), as well as compatible with measurements in
molecular clouds (Crutcher 1988, 1994, 1999).
On the other hand, the slope of the proton injection $p$ is determined also here by matching the corresponding slope for radio emission. The range of uncertainties does not seem to affect it significantly (see below).

Tabatabaei et al. (2007) confirms, in an study of M33, that non-thermal emission is spatially associated with star forming regions and propose that the higher the (synchrotron photon) energy the more important the contribution of localized structures (SNRs) may be. Although this maybe true, the spatial correlation by itself is not really a proof of it, since star forming regions also have an enhanced cosmic ray sea, and thus an enhanced production of secondary electrons which emit non-thermally. In fact Bressan, Silva and Granato (2002) have shown that the contribution of radio SNRs to the non-thermal emission cannot be dominant in galaxies, finding that in our Galaxy in particular, it cannot be more than about 6\%.  In any case, this is an uncertainty in determining the magnetic field in our model that must be emphasized: we are considering that most of the radio emission comes from cosmic ray electrons that have been either injected by SNRs (primaries) or other accelerators or produced in cosmic-ray interactions (secondaries from knock-on, charged pion decays). If there is a contribution to the total non-thermal radio signal made by unresolved SNRs, this would diminish the estimation of the average magnetic field in the starburst. This distinction cannot, however, be made with the data now at hand on M82.

Regarding the far IR emission, starbursts data generally requires a dust emissivity law, $\nu^{\sigma}B(E, T)$, where typically $\sigma = 1.5$ and $B(E, T)$ is the Planck function. This greybody peaks at $\sim$45 K and has a luminosity of $4 \times 10^{10} L_{\odot}$ (Telesco \& Harper, 1980, corrected by distance). However, at higher frequencies, simple greybody emission cannot explain observations, and an excess appears at near IR wavelengths.
Lindner (2003) proposed a dust cloud model in which an envelope of dust grains surrounds a luminous source. Each concentric shell decreases its density with increasing radius and has its own temperature and associated flux.
As a simpler but still accurate approximation, we show here that an addition of just one secondary greybody is enough to fit the data well for our aims, the temperature is warmer ($T\simeq$ 200 K) than the main one and has a smaller luminosity of $7 \times 10^9 L_{\odot}$. Small changes in the warm greybody luminosity or maximum of the peak are possible with results similar of the ones presented here, what we have explored. The fit to the IR data is quite good, as can be seen in Fig. \ref{Fig2}. \\

The model discussed in this
work yields inverse Compton fluxes of just a few percent or less than the
upper limits at X-ray energies (see e.g., Persic 2008 and references therein). 
The diffuse emission is overwhelmed by emission from compact objects.

 Fig. \ref{Fig5} shows the $\gamma$-ray differential flux that results from this model coming from the central region of M82, together with the sensitivity curves of possible instruments observing it. The goodness of this prediction is that it is consistent with the whole rest of the electromagnetic spectrum: the particles generating this emission, both hadrons and leptons, are the same ones generating the radiation seen at lower energies. The model confirms that there is not an expectation for detection by EGRET (consistent with EGRET upper limit by Torres et al. 2004, and the stacking analysis of EGRET data by Cillis et al. 2005) but predicts M82 as a source for Fermi. The total flux estimations at high and very high energies are as follows: 
for $E>100$ MeV, $2.6 \times 10^{-8}$ ($8.3 \times 10^{-9}$) cm$^{-2}$ s$^{-1}$;
for $E>100$ GeV, $8.8 \times 10^{-12}$ ($2.8 \times 10^{-12}$) cm$^{-2}$ s$^{-1}$, with the parenthesis representing the results obtained with the lower energy cutoff and lower supernova explosion rate.
Separate contributions are plotted from each $\gamma$-ray channel: neutral pion decay, bremsstrahlung and inverse Compton (against CMB, far and near IR photon densities).
Finally,  Fig. \ref{Fig5} also presents neutrino fluxes coming from the inner part of the starburst galaxy. The separate contribution of each neutrino process can be seen there, together with the total flux.  As in the previous case, parenthesis represent the results obtained with the lower energy cutoff and lower supernova explosion rate.
The total flux estimations at high and very high energies are as follows: 
for $E>100$ GeV, $1.2 \times 10^{-11}$ ($3.9 \times 10^{-12}$) cm$^{-2}$ s$^{-1}$; 
 for $E>1$ TeV $3.8 \times 10^{-13}$ ($9.9 \times 10^{-14}$) cm$^{-2}$ s$^{-1}$.

In the light of our results, only if the highest end of the predictions happens to be a realistic representation of the galaxy, the MAGIC telescope could detect it above 300 GeV with $5\sigma$ in 50h. Although the time required by MAGIC alone would be unrealistic in order to obtain a detection for the lowest end of the predictions, with the upcoming arrival of MAGIC II, the time to expend on this source could be affordable. The actual estimations for MAGIC II sensitivity are a factor of 2 to 3 better than for MAGIC I, so the flux coming from the starburst galaxy would be detected within 50h.
Similar estimations would apply for the VERITAS array. 
But indeed, the instrument of choice for this source would be the forthcoming Cerenkov Telescope Array (CTA, now in Design Study). Presenting an instrument acceptance extending both to the lower and higher energy ends of that of the previous telescopes, with a 1 TeV sensitivity one order of magnitude or more better than MAGIC I, M82 should, if this model is a realistic representation of it, be a bright source for CTA. With the latter, studies on the possible cutoff in the proton spectra could be made, although M82 would still look as a point like source. 

\begin{figure*} [t!]
\centering
\includegraphics[width=.48\columnwidth,trim=0 5 0 10]{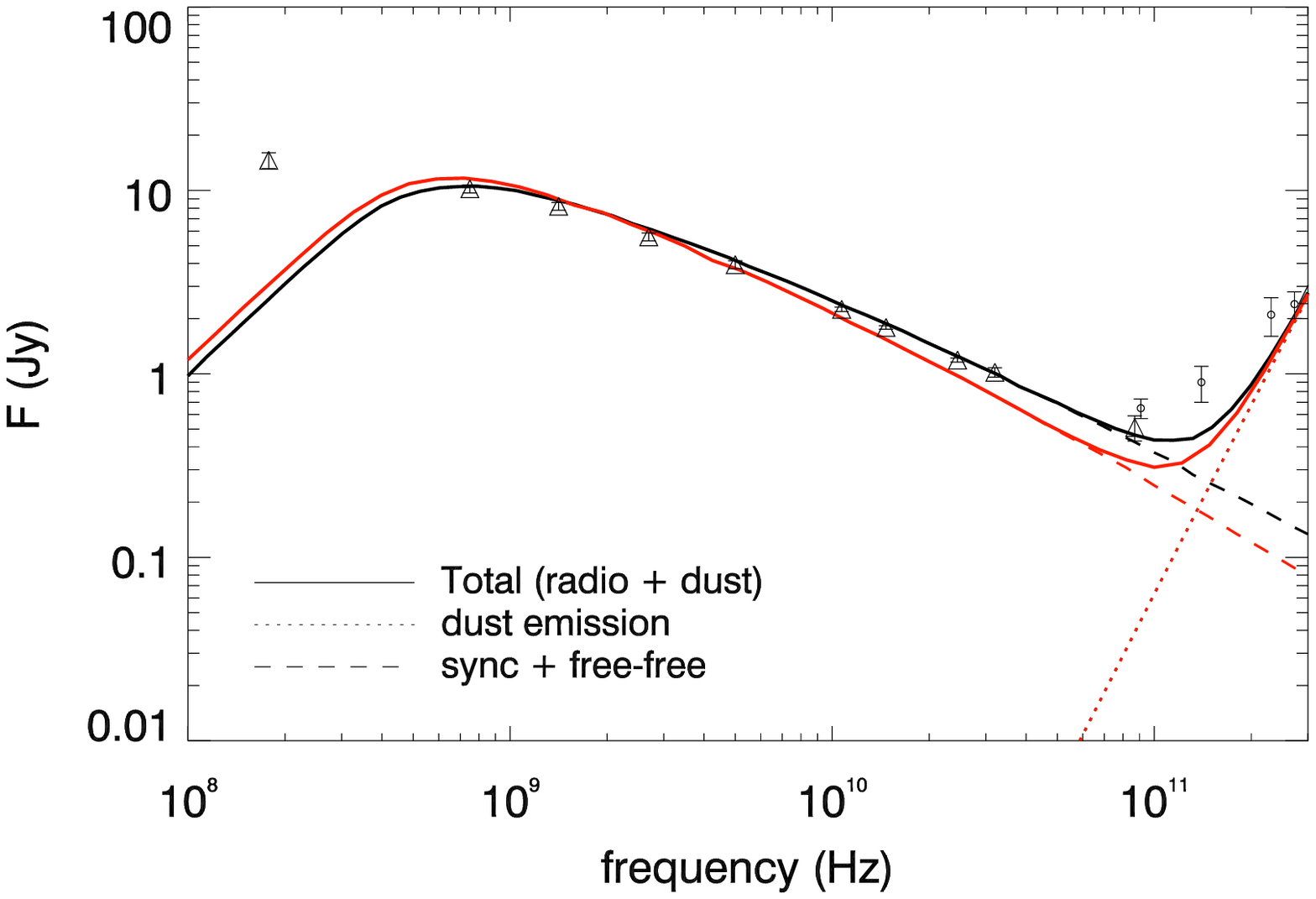}
\includegraphics[width=.48\columnwidth,trim=0 5 0 10]{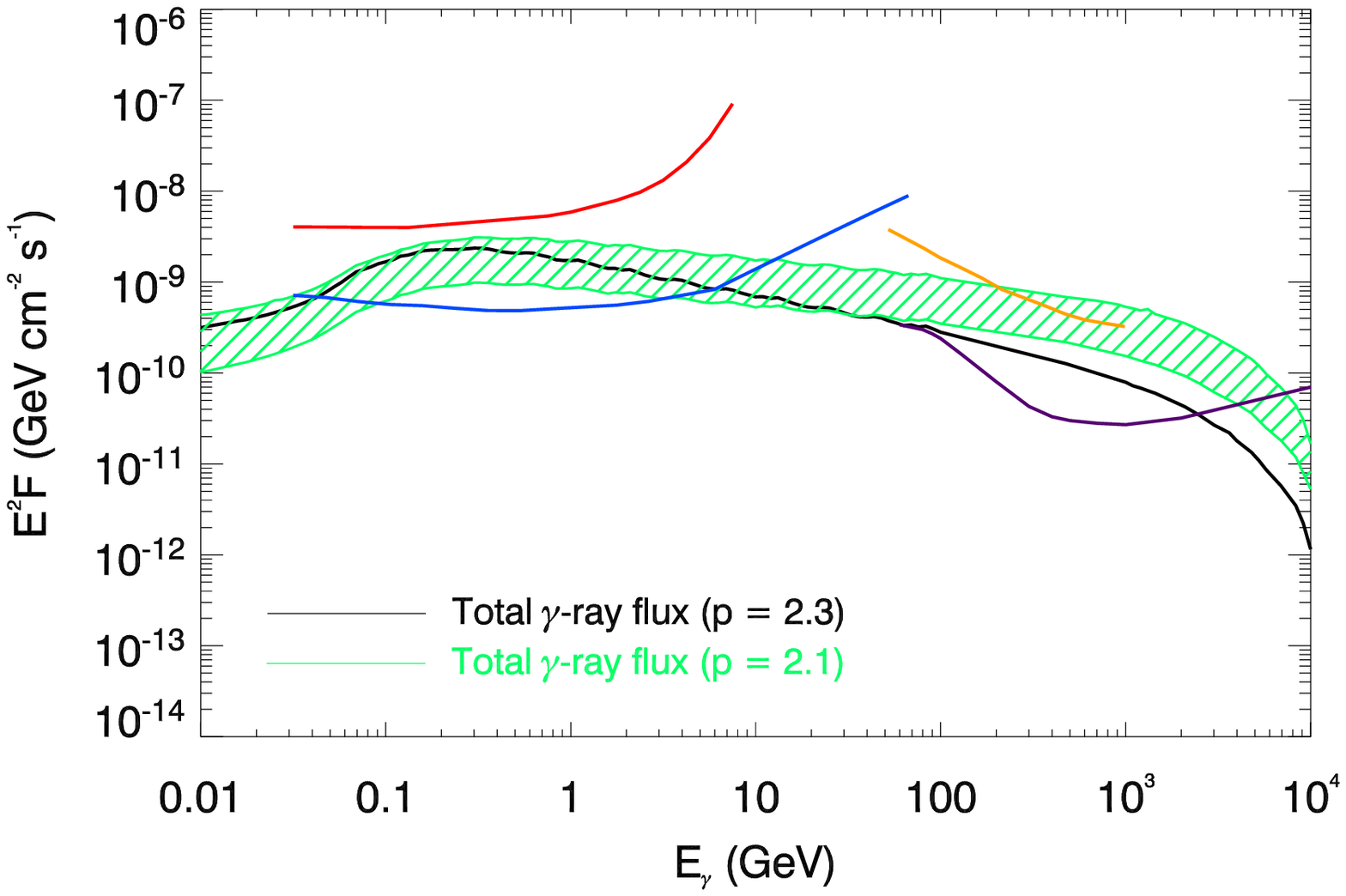}
\caption{Comparison of the multi-wavelength predictions for different initial spectral slope in the injection, see Fig. 3 for further details. The black curve correspond to our model with proton injection spectrum $p = 2.1$ and magnetic field $B = 130 \mu$G, whereas the red curve correspond to the results of modelling with $p = 2.3$ and $B = 170 \mu$G. The gamma-ray emission from the $-2.3$ model in the case of the highest SN explosion rate is shown against those obtained with the harder injection (the green shadow, coming from the uncertainties described above). Main differences of course appear at high energies.}
 \label{comp1}
\end{figure*}

We have explored the dependence of these results on the initial injection slope $p$ (enforcing it to be different) finding that, within a reasonable range, this is not relevant. In Fig. \ref{comp1} we show this by analyzing the predictions of our model for a $-2.3$ average injection slope. We show both the multi-wavelength predictions in radio-IR and the gamma-ray emission. We see that no relevant differences (within uncertainties) are found in the parameters that give rise to these curves.

We have mentioned that care should be exercised with equipartition estimates, since after the work by Beck \& Krause (2005), it was shown that the usual formula to compute the equipartition field (i.e.,  
or the minimum-energy estimate of total magnetic fields strengths from 
radio synchrotron intensities) maybe of limited practical use. It is based on the 
ratio $\cal K$ of the total energies of cosmic ray protons and electrons (which is generally computed without a full model of the system). In addition to other non-trivial technical problems mentioned by Beck \& Krause,  
if energy losses of electrons are important, the number density between these particles increases with particle energy
and the equipartition field may be underestimated significantly, the correct value can be computed only by
constructing a model of gas density and cosmic ray propagation.
Beck \& Krause (2005) already emphasize that starburst galaxies and regions of high star formation rate in the
central regions and massive spiral arms of galaxies have high gas densities
and mostly flat radio spectra, and non-thermal bremsstrahlung and other losses are important.   
Using the steady population of relativistic particles that we have found as a result of our model, we determine the energy content in such component, and from there we {\it impose} equipartition in order to see what would be the magnetic field determined in this way. It results in $150 \mu G$, which is close (but not the same) to the estimation by Weaver et al. (2002).
Further analysis of the details of the equipartition issue in starbursts will be presented elsewhere. \\

\begin{figure}[t]
\centering
\includegraphics[width=.5\columnwidth,trim=0 5 0 10]{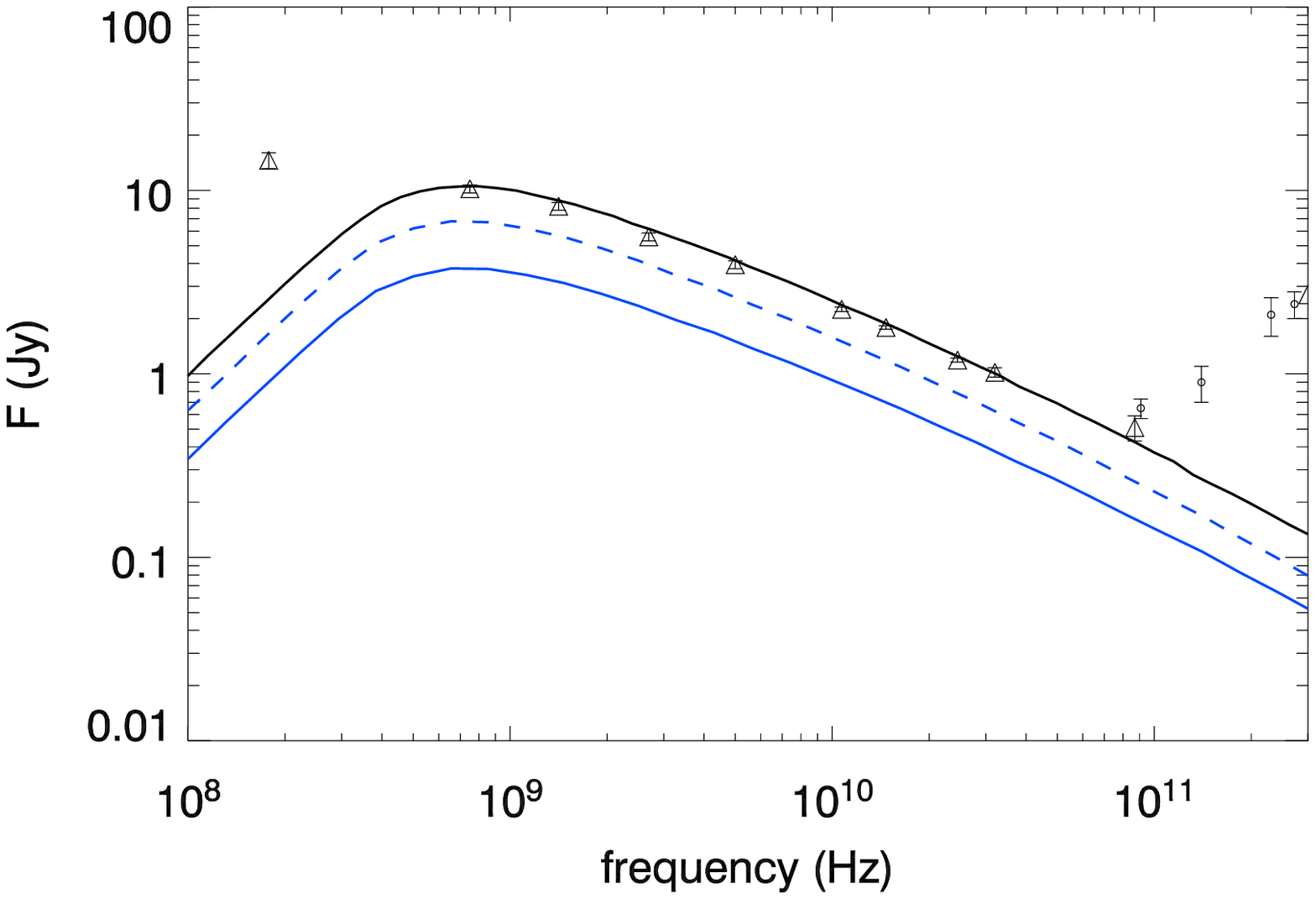}
\caption{Different contributions to the radio emission (synchrotron + free-free) by the steady primary-only (blue) and secondary-only (yellow) electron population, also compared to the total radio emission of the whole electron population (black).}
 \label{comp-radio}
\end{figure}

Loeb and Waxman (2006) have suggested that hadronic 
processes in starbursts can produce
a large enough background of diffuse high energy neutrinos to
be observable with ICECUBE. They arrive to this conclusion (criticized  by Stecker 2007) 
by assuming essentially two assumptions and making order-of-magnitude estimations thereafter: 
{\bf a)} that protons lose essentially
all of their energy to pion production, and {\bf b)} that a lower limit to the
energy loss rate of the protons can then be obtained from the synchrotron
radio flux by assuming that all of the electrons (and positrons) which
are radiating are from pion decay. Even when photo-dissociation in starbursts is a minor effect (Anchordoqui et al. 2008) assumption {\bf a} is arguable, as Stecker emphasized, 
because particles are subject to diffusion and convection by winds in addition to pion losses, i.e., the galaxy is not
a complete calorimeter, especially at the highest energies. 
Assumption {\bf b} is arguable since 
synchrotron radiating
electrons are not only the secondaries, but the primaries too. Fig. \ref{comp-radio} quantifies this for M82 and compares the contribution to
the synchrotron radiation of the inner starburst if only-primary or only-secondary electrons are considered. From about $E_e-m_e \sim 10^{-1}$ to $\sim 100$ GeV, the secondary population of electrons slightly dominates, see Fig. \ref{comp}, left, with the primary population contributing significantly. It would then be an error to normalize the whole radio emission to the secondaries alone (or primaries alone) and then use this to fix the energy loss rate of protons, from where to estimate the neutrino emission. As far as we are aware, this paper presents the first full
neutrino emission computation from a starburst galaxy which is self-consistent with the emission at all electromagnetic frequencies as part of the same model. Given the detail needed for such a description, a generalization of our M82 results to all starbursts is not recommended. We remark, however,  that if this is done, and if we support it by basic assumptions on the starbursts number density within the horizon, the neutrino diffuse emission would be below the Waxman-Bahcall limit.

The 
sky-average sensitivity of the  ICECUBE installation (IC-22, with 22 strings) at 90\%C.L. to a generic 
$E^{-2}$ flux of $\nu_\mu$ is $ 1.3 (2.0) \times 10^{-11}$ 
TeV$^{-1}$ 
cm$^{-2}$ 
s$^{-1}$ 
$(E/{\rm TeV})^{-2}$ depending on the analysis (Bazo Alba et al. 2008). This is not enough to detect M82 directly (see Fig. \ref{Fig5}). In fact, using our estimation of the neutrino flux from M82 and following the comments made by Anchordoqui et al. (2004), we find than $<2$ event per year would be expected in the full ICECUBE facility,  and a more definitive assessment of its sensitivity to 
such a signal will need to await further refinement of angular and 
energy resolutions, via improved knowledge of the detector response. \\

We acknowledge  support by grants
AYA 2006-00530, CSIC-PIE 200750I029, and AYA2008-01181-E/ESP.
The work of E. de Cea del Pozo has been made under the auspice of a FPI 
Fellowship, grant BES-2007-15131. We acknowledge Luis Anchordoqui for discussions.\\


\begin{thebibliography}{}



\bibitem{} Aky\"uz, A., Brouillet, A. \& \"Ozel, M.E. 1991, A\&A, 248, 419

\bibitem{} Anchordoqui  L.A., Goldberg H., Halzen F. \& Weiler T. J. 2004, Phys. Lett. B600, 202

\bibitem{} Anchordoqui  L.A., Hooper D.,  Sarkar S.,  \& Taylor A. M. 2008, Astropart. Phys. 29, 1 

\bibitem{} Bazo Alba J. L. et al. 2008, arXiv:0811.4110


\bibitem{} Bartel N. et al. 1987, ApJ, 322, 505
\bibitem{} Beck R., \& Krause M. 2005, AN 326, 414
\bibitem{} Bell A. R. 1978, MNRAS 182, 443 
\bibitem{} Berezhko E. G., Ksenofontov L. T., V{\"o}lk H. J. 2006, A\&A 452, 217

\bibitem{} Blom J. J., Paglione T. A. \& Carraminana 1999, 516, 744
\bibitem{} Bressan A., Silva L. \& Granato G. L. 2002, A\&A 392, 377 
\bibitem{} Brown R. L. \& Marscher A. P. 1977, ApJ 212, 659 

\bibitem{} Carlstrom, J.E. \& Kronberg, P.P. 1991, ApJ, 366, 422
\bibitem{} Casasola, V., Bettoni, D. \& Galletta, G. 2004, A\&A, 422, 941

\bibitem{} Cillis A. N., Torres D. F. \& Reimer O. 2005,   ApJ {621}, 139
\bibitem{} Crutcher, R. M., Rogstad, D. H. \& Chu, K. 1978, ApJ, 225, 784

\bibitem{} Crutcher R. M. 1988, in ``Molecular Clouds, Milky-Way \&
External Galaxies", R. Dickman, R. Snell, \& J. Young Eds., New
York,  Springer, p.105

\bibitem{} Crutcher R. M. 1994,  ``Clouds, cores and low mass stars",
Astronomical Society of the Pacific Conference Series, volume 65;
Proceedings of the 4th Haystack Observatory, edited by D. P.
Clemens and R. Barvainis, p.87

\bibitem{} Crutcher R. M. 1999, ApJ 520, 706


\bibitem{} de Grijs R., O'Connell R.W. \& Gallagher J.S. 2001, AJ, 121, 768
\bibitem{} Drury L. O. C., Aharonian F. A. \& Voelk H. J. 1994, A\&A 287, 959
\bibitem{} Domingo-Santamar\'{\i}a E. \& Torres D. F. 2005, A\&A, 444, 403
\bibitem{} Everett J. E. et al. 2008, ApJ 674

\bibitem{} Fatuzzo M. \& Melia F. 2003, ApJ 596, 1035 
\bibitem{} Fazio G. G. 1967, ARA\&A 5, 481 
\bibitem{} F\"orster, N.M. et al. 2003,A\&A 399, 833
\bibitem{} Freedman, W.L., et al. 1994, ApJ, 427, 628
\bibitem{} Funk S., Reimer O., Torres D. F. \& Hinton J. A. 2008, ApJ 679, 1299
 
 \bibitem{} Gallagher J. S. \& Smith L. J. 2005, Extra-planar Gas 
ASP Conference Series, Vol. 331, 2005 
Robert Braun Editor
\bibitem{} Ginzburg V. L. \& Syrovatskii S. I. 1964, ``The origin of
cosmic rays", Pergamon Press, Oxford, England.

\bibitem{} Hughes, D.H., Gear, W.K. \& Robson, E.I. 1994, MNRAS, 270, 641

\bibitem{} Kelner S. R., Aharonian F. A. \& Bugayov V. V. 
 2006, Phys. Rev. D74, 034018
\bibitem{} Klein, U., Wielebinski, R. \& Morsi, H.W. 1988, A\&A, 190, 41
\bibitem{} Kronberg, P.P. \& Wilkinson, P.N. 1975, ApJ, 200 430
\bibitem{} Kronberg, P.P., Biermann, P. \& Schwab, F.R. 1985, ApJ, 291, 693
\bibitem{} Kronberg, P.P. \& Sramek, R.A. 1985, Science 227, 28
\bibitem{} Lindner, J. 2003, JCEP, 2, 1

\bibitem{} Longair M. S. 1994, ``High Energy Astrophysics, Vol.2:
Stars, the Galaxy and the Interstellar Medium", Cambridge
University Press, 2nd 
ed.

\bibitem{} Loeb, A. and Waxman, E. 2006 JCAP 05, 003
\bibitem{} Mannheim K. \& Schlickeiser R. 1994, A\&A 286, 983
\bibitem{} Maraschi L., Perola G. C. \& Schwarz S. 1968, Il Nuovo Cimento LIII B, 1975 
\bibitem{} Markoff S., Melia, F. \& Sarcevic I. 1999, ApJ 522, 870 
\bibitem{} Marscher A. P. \& Brown R. L. 1978, ApJ 221, 588 


\bibitem{} Mayya, Y.D., Carrasco L. \& Luna, A. 2005, ApJ, 628, L33
\bibitem{} Mayya, Y.D., Bressan, A., Carrasco, L. \& Hernandez-Martinez, L. 2006, ApJ, 649, 172
\bibitem{} McLeod, K.K., Rieke, G.H., Rieke, M.J. \& Kelly, D.M. 1993, ApJ, 412, 111
\bibitem{} Moskalenko I. \& Strong A. 1998, ApJ 493, 694

\bibitem{} Paglione T. A. D., Marscher A. P., Jackson J. M. \& Bertsch D. L. 1996, ApJ 460, 295
\bibitem{} Pavlidou, V., \& Fields, B. 2001, ApJ, 558, 63
\bibitem{} Persic, M., Rephaeli, Y. \& Arieli, Y. 2008,  A\&A, 486, 143
\bibitem{} O'Connell, R.W. \& Mangano, J.J. 1978, ApJ, 221, 62  
\bibitem{} Orth C. D. \& Buffington A. 1976, ApJ 206, 312
\bibitem{} Ramaty R. \& Lingenfelter R. E. 1966, Journal of Geophysical Research 71, 3687 
\bibitem{} Ranalli, P., Comastri, A., Origlia, L. \& Maiolino, R. 2008, MNRAS, 386, 1464
\bibitem{} Rieke, G.H., et al. 1980, ApJ, 238, 24
\bibitem{} Sakai, S. \& Madore, B.F. 1999, ApJ, 526, 599
\bibitem{} Stecker F. W. 1969, Astrophysics and Space Science 6, 377
\bibitem{} Stecker F. W. 1977, ApJ 212, 60 
\bibitem{} Stecker F. 2007, J. Phys. Conf. Ser. 60, 215 
\bibitem{} Strickland, D.K., Ponman, T.J. \& Stevens, I.R. 1997, A\&A, 320, 378
\bibitem{} Strong A. \& Moskalenko I. 1998, ApJ 509, 212

\bibitem{} Tabatabaei F. S. et al. 2007, A\&A 475, 133
\bibitem{} Telesco, C.M. \& Harper, D.A. 1980, ApJ, 235, 392
\bibitem{} Telesco, C. M., Joy, M., Dietz, K., Decher, R. \& Campins, H. 1991, ApJ, 369, 135 
\bibitem{} Torres D. F. 2004, ApJ 617, 966
\bibitem{}Torres D. F., Reimer O., Domingo-Santamar\'{\i}a E. \& Digel S.
2004, ApJ 607, L99
\bibitem{}Torres D. F., Domingo-Santamar\'{\i}a E. 2005 Mod. Phys. Lett. A20, 2827
\bibitem{} V\"olk, H.J., Aharonian, F.A. \& Breitschwerdt, D. 1996, SSRv, 75, 279
\bibitem{} V\"olk, H.J., Klein, U., \& Wielebinski, R. 1989, A\&A, 213, L12

\bibitem{} Weaver K. A., Heckman T. M., Strickland D. K. \& Dahlem
M. 2002,  ApJ 576, L19
\bibitem{} Weiss, A., Neininger, N., H\"uttemeister, S. \& Klein, U. 2001, A\&A, 365, 571
\bibitem{} Young, J. S. \& Scoville, N. Z. 1984, ApJ, 287, 153

\end{thebibliography}
\end{document}